\documentstyle[12pt]{article}
\renewcommand {\thefootnote}{\fnsymbol{footnote}}
\begin{document}
\thispagestyle{empty}
\large
\renewcommand {\thefootnote}{\fnsymbol{footnote}}
\renewcommand{\thesection}{\Roman{section}}
\def \beq {\begin{equation}}
\def \eeq {\end{equation}}
\def \bes {\begin{eqnarray}}
\def \ees {\end{eqnarray}}
\def\rv{\mbox{\boldmath$r$}}
\def\drv{{d\mbox{\boldmath$r$}}}
\def\mum {\,\mu\mbox{m}}
\begin{center}
{\bf Complete roughness and conductivity corrections
\\
for the recent Casimir force measurement}
\\[5mm]
G.~L.~Klimchitskaya,$\!{}^{1,}$\footnote{On leave from 
North-West Polytechnical Institute,
St.Petersburg, Russia.
Electronic address: galina@GK1372.spb.edu}
Anushree Roy,${}^2$
U.~Mohideen,$\!{}^{2,}$\footnote{Electronic address:
umar.mohideen@ucr.edu}
and V.~M.~Mostepanenko$\!{}^{1,}$\footnote{On 
leave from
A.Friedmann Laboratory for Theoretical Physics,
St.Petersburg, Russia. Electronic address:
mostep@fisica.ufpb.br}
\\[0.5cm]
${}^1\,${\it Physics Department, Federal University
of Para\'{\i}ba, C.P.5008, \\
CEP 58059--970, Jo\~{a}o Pessoa, Pb---Brazil}
\\[3mm]
${}^2\,${\it Department of Physics, University of
California, Riverside,
\\California 92521}
\end{center}

\begin{abstract}
We consider detailed roughness and conductivity
corrections to the Casimir force in the recent
Casimir force measurement employing an atomic force
microscope. The roughness of the test bodies ---
a metal plate and a sphere --- was investigated with
the AFM and the SEM respectively.
It consists of separate crystals of different heights
and a stochastic background. The amplitude of roughness
relative to the zero roughness level was determined and
the corrections to the Casimir force were calculated
up to the fourth order in a small parameter (which is
this amplitude divided by the distance between the two
test bodies). Also the corrections due to finite
conductivity were found up to the fourth
order in relative penetration depth of electromagnetic
zero point oscillations into the metal. 
The theoretical result for the configuration of a sphere
above a plate taking into account both corrections
is in excellent agreement with the measured
Casimir force.
\end{abstract}

\section{INTRODUCTION}

The Casimir effect \cite{Cas} which arises in bounded
regions and in spaces with non-trivial topology
is of great interest to specialists in the most
diverse fields of physics --- from statistical and
atomic physics to elementary particle physics and
cosmology. It explores the dependence of the 
vacuum polarization on the geometrical
parameters of the quantization domain, leading to attractive and
repulsive forces acting between the boundaries
(see the review papers \cite{P-M-G, M-T-r} and the
monographs \cite{Mil, book}).

A considerable amount of recent attention has been focussed on
experimental verification of the Casimir force law
between metallic surfaces. The first experiment
of this kind was performed more than forty years ago
\cite{Sparn} and provided qualitative
confirmation of the Casimir prediction. Then over
a period of years the force between dielectric test bodies was
used to measure the Casimir force (see, e.g.,
\cite{T-W, Hunk} and the other references in
\cite{Mil,book}). During this period only one paper may
be cited \cite{VB-O} where the Casimir force between
the plate and the spherical lens covered by
Chromium layers was measured. It should be noted that 
Chromium is a poor reflector for a large portion of the 
measured distances.  In this paper,
considerable attention has been given to the finite
conductivity corrections to the Casimir force. Also
the possible corrections due to surface roughness were
discussed qualitatively. In all the earlier experiments only
variants of the spring balance were used to measure
the force.

In the paper \cite{Lam} which opens the modern stage
of the Casimir force measurements between metals 
the distance range from $0.6\mum$ to $6\mum$ was
investigated. The test bodies were $Cu$ plus $Au$
coated quartz optical flat, and a spherical lens.
The torsion pendulum was used to measure the Casimir force.
As mentioned in \cite{Lam}, experimental data does
not support the presence of finite conductivity
corrections which are negative and can achieve
20\% of the net Casimir force at the closest spacing.
The roughness corrections which can achieve 20--30\%
of the net result if there are deviations of 
the interacting surfaces from the perfect shape
\cite{K-P} were not investigated in \cite{Lam}.
As discussed in \cite{Lam} also, the data is not of
sufficient accuracy to demonstrate the finite
temperature corrections. We would like to remind that
the temperature correction at room temperature
is of 129\% and 174\% of
the net force when the space separation is
correspondingly $5\mum$ and $6\mum$.
The values of both the Casimir force and temperature
correction to it at such distances are of the order
of $10^{-12}\,$N. Their experimental measurement
and investigation is the unresolved problem of
paramount importance.

In the paper \cite{M-R} an Atomic Force Microscope (AFM)
was used to make a precision measurement of the
Casimir force between a sphere and a flat plate
covered by the $Al$ and $Au/Pd$ layers.
The measurements were done for plate-sphere
separations between $0.1\mum$ to $0.9\mum$.
The experimental data was shown to be consistent
with the theoretical calculations including the
finite conductivity and roughness corrections
 calculated up to the second order in
appropriate parameters \cite{B-K-R}. 
No account has been taken in
these calculations of the specific shape of roughness
peculiar to the test bodies in use.  Also the third
and fourth orders of these corrections were neglected
although they could contribute to the
comparison of the theory and experiment at an
accuracy level of 1\% for the smallest separations
(temperature corrections are not important in this
distance range).

Other techniques for measuring the Casimir force
have also been proposed (see, e.g., \cite{Carug, Grado}).

Here we present the complete experimental and
theoretical investigation of the surface roughness
and roughness corrections to the Casimir force in
the experiment \cite{M-R}. For this purpose the
roughness of the plate was measured with the 
AFM and the roughness of the sphere --- by the 
Scanning Electron Microscope (SEM). The
surface is composed of large separate crystals
situated irregularly on the surface. They are modeled
by parallelepipeds of two different 
heights situated on the stochastic background.
The corresponding corrections to the Casimir force are
computed by the use of the approximate method
proposed earlier in \cite{M-S,B-K-M}. The corrections
due to roughness up to the fourth order in relative
roughness amplitude are obtained.

To provide the higher order finite conductivity
corrections the measurement range is subdivided into
ranges of small and large distances. It is shown
that at small distances it is possible to neglect 
the external $Au/Pd$ cap layer. At large distances
the effective penetration depth of the electromagnetic
zero point oscillations into the metal is found. As a
result the corrections due to finite conductivity
up to the fourth order are calculated taking into account
the effect of the surface roughness.

The resulting Casimir force with both corrections is
in excellent agreement with the experimental data.

The paper is organized as follows. In Sec.~II the
necessary details of the experiment \cite{M-R} are
reviewed.
Sec.~III contains a brief formulation of the perturbative
approach to the calculation of roughness corrections.
In Sec.~IV the investigation of surface roughness
and the roughness corrections is presented in relation to 
the experiment \cite{M-R}. Sec.~V
is devoted to the corrections due to the finite
conductivity of the metals. Here, the final expressions
for the Casimir force including both corrections are 
also obtained. In Sec.~VI they are compared with the
experimental data of \cite{M-R}. Sec.~VII contains
conclusions and discussion.

Throughout the paper units in which 
$\hbar=c=1$ are used.

\section{THE MEASUREMENT OF THE CASIMIR FORCE}

In Ref. [12] a standard AFM was used to measure the force between a
metallized sphere and flat plate at a pressure of 50\,mTorr
 and at room temperature. Polystyrene
spheres of $200\pm4\mum$ diameter were mounted on
the tip of $300\mum$ long cantilevers with $Ag$
epoxy. A 1.25\,cm diameter optically polished
sapphire disk was used as the plate. The cantilever
(with sphere) and plate were then coated with
300\,nm of $Al$ in an evaporator. Aluminum is used because of 
its high reflectivity for wavelengths
(sphere-plate separations)$\> >100\,$nm. Both surfaces
were then coated with less than 20\,nm layer of
$60\%Au/40\%Pd$. The sphere diameter was measured
using the SEM to be
$196.0\pm0.5\mum$.

In the AFM, the force on a cantilever is measured 
by the deflection of its tip. A laser beam is 
reflected off the cantilever tip to measure its
deflection. A force on the sphere would result in
a cantilever deflection leading to a difference signal
between photodiodes A and B (shown in Fig.~1). This
force and the corresponding cantilever deflection
are related by Hooke's law: $F=k\Delta z$, where $k$
is the force constant and $\Delta z$ is the 
cantilever deflection. The piezo extension with
applied voltage was calibrated with height standards
and its hysteresis was measured. The corrections due
to the piezo hysteresis (2\% linear correction) and
cantilever deflection (discussed in \cite{M-R}) were
applied to the sphere-plate separations in all
collected data.

To measure the Casimir force between the sphere and 
the plate they are grounded together with the AFM.
The plate is then moved towards the sphere in 3.6\,nm steps 
and the corresponding photodiode difference
signal was measured (approach curve). The signal
obtained for a typical scan is shown in Fig.~2.
Here ``0" separation stands for contact of the sphere
and plate surfaces. It does not take into account the
absolute average separation between the $Au/Pd$
layers due to the surface roughness which is about
80\,nm (see Sec.~IV). If one also takes into account  
the $Au/Pd$ cap layers which are transparent at small
separations (see Sec.~V) the absolute 
average separation at contact between $Al$ layers is
about 120\,nm. Region 1 shows that the force curve 
at large separations is dominated by a linear signal.
This is due to increased coupling of scattered light
into the diodes from the approaching flat surface.
Embedded in the signal is a long range attractive
electrostatic force from the contact potential
difference between the sphere and plate, and the
Casimir force (small at such large distances).
In region 2 (absolute separations vary from contact
to 350\,nm) the Casimir force is the dominant
characteristic far exceeding all the systematic
errors. Region 3 is 
the flexing of the cantilever resulting from the
continued extension of the piezo after contact of the
two surfaces. Given the distance moved by the flat
plate ($x$-axis), the difference signal of the
photodiodes can be calibrated to a cantilever
deflection in nanometers using the slope of the curve
in region 3.

Next, the force constant of the cantilever was
calibrated by an electrostatic measurement. The sphere
was grounded to the AFM and different voltages in the
range $\pm0.5\,$V to $\pm3\,$V were applied to the
plate. The force between a charged sphere and plate
is given as \cite{elect}
\beq
F=2\pi\epsilon_0(V_1-V_2)^2
\sum\limits_{n=1}^{\infty}
\mbox{csch}n\alpha(\coth\alpha-n\coth n\alpha).
\label{u1}
\eeq
\noindent
Here $V_1$ is the applied voltage on the plate, $V_2$
represents the residual potential on the grounded
sphere, and $\epsilon_0$ is the permittivity of free
space. One more notation is 
$\alpha=\cosh^{-1}(1+a/R)$, where $R$ is the radius
of the sphere and $a$ is the separation between the
sphere and the plate. From the difference in force
for voltages $\pm V_1$ applied to the plate, we can
measure the residual potential on the grounded
sphere $V_2$ as 29\,mV. This residual potential is
a contact potential that arises from the different
materials used to ground the sphere. The electrostatic
force measurement was repeated at 5 different
separations and for 8 different voltages $V_1$.
Using Hooke's law and the force from Eq.\,(\ref{u1}),
we measure the force constant of the cantilever $k$.
The average of all the measured $k$ is 0.0182\,N/m.

The systematic error corrections to the force curve
of Fig.~2, due to the residual potential on the
sphere and the true separations between the two
surfaces, are now calculated. Here the near linear
force curve in region 1, is fit to a function of the
form: $F=F_c(a+a_0)+B/(a+a_0)+C\times(a+a_0)+E$.
Here $a_0$ is the absolute separation at contact,
which is constrained to $120\pm5\,$nm, is the only
unknown to be completely obtained by the fit. The
second term represents the inverse linear dependence
of the electrostatic force between the sphere and
plate for $R\gg a$ as given by Eq.\,(\ref{u1}).
The constant $B=-2.8\,\mbox{nN}\cdot$nm corresponding
to $V_2=29\,$mV and $V_1=0$ in Eq.\,(\ref{u1}) is
used. The third term represents the linearly increasing
coupling of the scattered light into the photodiodes
and $E$ is the offset of the curve. Both $C$ and $E$
can be estimated from the force curve at large
separations. The best fit values of $C$, $E$ and
the absolute space separation $a_0$ are determined by
minimizing the $\chi^2$. The finite conductivity
correction and roughness correction (the largest
corrections) do not play a significant role in the
region 1 (see Sec.~VI) and thus the value of $a_0$
determined by the fitting is unbiased with respect
to these corrections. These values of $C$, $E$ and
$a_0$ are then used to subtract the systematic errors
from the force curve in region 1 and 2 to obtain the
measured Casimir force as $(F_c)_m=F_m-B/a-Ca-E$,
where $F_m$ is the measured total force.

This procedure is repeated for 26 scans in different
locations of the flat plate. The average measured
Casimir force $(F_c)_m$ as a function of sphere-plate
separations from all the scans is shown in Figs.~4,5
below as open squares.
 
\section{ROUGHNESS CORRECTIONS TO THE CASIMIR FORCE}
    
For  distances of $a\sim 1\mum$ between the
interacting bodies the surface roughness makes
an important contribution
 to the value of the Casimir force.
Although the exact calculation of roughness
contribution is impossible, one can find the
corresponding corrections approximately with the
required accuracy.
In the case of stochastic roughness the corrections to
the van der Waals and Casimir forces were first
calculated in \cite{Bree} up to the second order in
relative roughness dispersions (the fourth order corrections
were obtained in \cite{BKM1}). Effects of large-scale surface
roughness on only the non-retarded van der Waals force were
investigated in \cite{M-Maz,Maz-M}.

 The method of greatest practical
utility is the summation of retarded interatomic
potentials over all atoms of two bodies distorted by
roughness with a subsequent multiplicative
normalization of the interaction coefficient
\cite{M-S,B-K-M}
\beq
U(a)=-\frac{CN_1N_2}{K}
\int\limits_{V_1}\!\drv_1
\int\limits_{V_2}\!\drv_2
|\rv_1-\rv_2|^{-7}.
\label{t1}
\eeq
\noindent
Here $N_{1,2}$ are the numbers of atoms per unit
volume of the bodies, $C$ is the constant of the
retarded van der Waals interaction, $K$ is a special
normalization constant, $a$ is a distance between
bodies. 

The appropriate choice of the normalization
constant $K$ gives the possibility of increasing the
accuracy of additive summation. Its value can be
found as a ratio of the additive and exact potentials
for the configuration admitting the exact solution.
For two plane parallel plates, as an example most
important for experiment,
\beq
K=\frac{CN_1N_2}{\Psi(\varepsilon_1,\varepsilon_2)}
>1,
\label{t2}
\eeq
\noindent 
where the 
$\,\Psi$ is defined as \cite{L-P}
\bes
&&\Psi(\varepsilon_1,\varepsilon_2)=
\frac{5}{16\pi^3}
\int\limits_{0}^{\infty}
\int\limits_{1}^{\infty}
\frac{x^3}{p^2}\left\{\left[
\frac{(s_1+p)\,(s_2+p)}{(s_1-p)\,(s_2-p)}\,e^x-1
\right]^{-1}\right.
\nonumber \\
&&\phantom{aaaaaaaaaaa}+
\left.\left[
\frac{(s_1+p\varepsilon_1)\,(s_2+
p\varepsilon_2)}{(s_1-p\varepsilon_1)\,
(s_2-p\varepsilon_2)}\,e^x-1
\right]^{-1}\right\}dpdx.
\label{t3}
\ees
\noindent
Here, $s_{1,2}=(\varepsilon_{1,2}-1+p^2)^{1/2}$,
and $\varepsilon_{1,2}$ are
the static dielectric permittivities of the plate
materials.

From (\ref{t1}) and (\ref{t2}) the Casimir force is
\beq
F=-\frac{\partial U}{\partial a},
\qquad
U(a)=-\Psi(\varepsilon_1,\varepsilon_2)
\int\limits_{V_1}\!\drv_1
\int\limits_{V_2}\!\drv_2
|\rv_1-\rv_2|^{-7}.
\label{t4}
\eeq

For the configuration of two plane parallel plates,
 Eq.\,(\ref{t4}) is exact by construction. 
We use here the word ``exact" implying that the
approximative method does not 
bring any additional error.
Actually, the so called ``exact" results are
obtained in the approximation of large distances with
the proviso that $a\gg\lambda_0$, where $\lambda_0$
is the characteristic wavelength of absorption
spectra. At the same time the values of $a$ must
satisfy the condition $aT\ll 1$, where $T$ is a
temperature measured in energy units.
For two plates or a plate and 
a lens or a sphere of large curvature
radius covered by roughness the relative error of
the results obtained by (\ref{t4}) does not
exceed $10^{-2}\,$\% \cite{B-K-M} (this is proved
under the supposition that the roughness amplitude
$A$ is much smaller than $a$). Because of this 
the proposed method is very useful for the
calculation of roughness contribution in experiments
on the Casimir force.

Recently, other methods for approximative
calculation of the Casimir force have been proposed. Among
them the semiclassical \cite{S-S} and macroscopic
\cite{F} approaches
are applicable to the case of a sphere
near a wall. They do not take into
account the surface roughness. Also the path-integral
approach was suggested \cite{G-K} to study the space
and time deformations of the perfectly reflecting
boundaries. It was applied to describe the model
example of corrugated plates when the lateral
component of the Casimir force arises.

Now consider a plane plate (disk) of dimension $2L$,
thickness $D$ and a sphere above it of
radius $R$ both covered by roughness.
The roughness on the plate is described by the
function
\beq
z_1^{(s)}=A_1 f_1(x_1,y_1),
\label{t5}
\eeq
\noindent
where the value of amplitude is chosen in such a way
that $\max|f_1(x_1,y_1)|=1$. It is suitable to fix the
zero point in the $z$ axis by the condition
\beq
\langle z_1^{(s)}\rangle=
A_1\langle f_1(x_1,y_1)\rangle\equiv
\frac{A_1}{4L^2}
\int\limits_{-L}^{L}\!dx_1
\int\limits_{-L}^{L}\!dy_1f_1(x_1,y_1)=0.
\label{t6}
\eeq

The roughness on the sphere is most conveniently
described in the polar coordinates
\beq
z_2^{(s)}=a+R-\sqrt{R^2-\rho^2}+A_2f_2(\rho,\varphi).
\label{t7}
\eeq
\noindent
The value of the amplitude is chosen as specified
above. The value of $R$ in Eq.\,(\ref{t7}) is
defined in such a way that 
$\langle f_2(\rho,\varphi)\rangle=0$.

The potential $U$ from Eq.\,(\ref{t4}) for
configuration of a plate and a sphere with roughness
described by (\ref{t5}), (\ref{t7}) can be
represented as
\beq
U(a)=-\Psi(\varepsilon_1,\varepsilon_2)
\int\limits_{0}^{2\pi}\!d\varphi
\int\limits_{0}^{R}\!\rho d\rho
\int\limits_{z_2^{(s)}}^{a+2R}\!dz_2
U_A(\rho,\varphi,z_2),
\label{t8}
\eeq
\noindent
where
\beq
U_A(\rho,\varphi,z_2)=\!
\int\limits_{-L}^{L}\!\!dx_1\!
\int\limits_{-L}^{L}\!\!dy_1\!\!
\int\limits_{-D}^{z_1^{(s)}}\!
\frac{dz_1}{\left[(x_1-\rho\sin\varphi)^2+
(y_1-\rho\cos\varphi)^2+
(z_1-z_2)^2\right]^{\frac{7}{2}}}.
\label{t9}
\eeq
\noindent

In Ref.\cite{K-P} the perturbation theory was
developed in small parameters $A_{1,2}/a$ based on
Eqs.\,(\ref{t4}), (\ref{t8}), (\ref{t9}). All 
the results were obtained in the zeroth order of
the parameters $a/D$, $a/L$, and $a/R$ which
are much smaller than $A_{1,2}/a$ (in Ref.\cite{B-K-R}
it was shown that the corrections due to the
finiteness of a plate are negligible).
The perturbation expansion for the Casimir force is
\beq
F_R(a)=F_0(a)\sum\limits_{k=0}^{4}
\sum\limits_{l=0}^{4-k}
C_{kl}\left(\frac{A_1}{a}\right)^k
\left(\frac{A_2}{a}\right)^l,
\label{t10}
\eeq
\noindent
where the force acting between the perfect plate and
the sphere is
\beq
F_{0}(a)=-\Psi(\varepsilon_1,\varepsilon_2)
\frac{\pi^2R}{15a^3}.
\label{t11}
\eeq
\noindent
When the plate and the sphere are perfect metals
we have the limiting case 
$\varepsilon_{1,2}\to\infty$,
$\Psi\to\pi/24$ and Eq.\,(\ref{t11}) takes the form
\beq
F_{0}(a)=-
\frac{\pi^3R}{360a^3}.
\label{t11a}
\eeq

The first coefficient of Eq.~(\ref{t10}) is 
$C_{00}=1$. The other coefficients were found in
Ref.\cite{K-P} for the configuration of a lens 
(sphere) above
a plate and in Ref.\cite{B-K-M} for two plane
parallel plates. They are complicated integrals
involving functions describing roughness. In the 
case that
\beq
d_p,d_s\ll\sqrt{aR},
\label{t12}
\eeq
\noindent
where $d_p,d_s$ are the characteristic lateral sizes of
distortions covering the plate and the sphere, the
simple universal expression for the expansion
coefficients of Eq.(\ref{t10}) can be obtained.
As a result, Eq.(\ref{t10}) takes the form
\bes
&&F_R(a)=F_0(a)\left\{1+
6\!\left[
\langle\!\langle f_1^2\rangle\!\rangle
\left(\frac{A_1}{a}\right)^2-
2\langle\!\langle f_1f_2\rangle\!\rangle
\frac{A_1}{a}
\frac{A_2}{a}+
\langle\!\langle f_2^2\rangle\!\rangle
\left(\frac{A_2}{a}\right)^2
\right]\right.
\nonumber\\
&&\phantom{a}+
10\left[
\langle\!\langle f_1^3\rangle\!\rangle
\left(\frac{A_1}{a}\right)^3-
3\langle\!\langle f_1^2f_2\rangle\!\rangle
\left(\frac{A_1}{a}\right)^2
\frac{A_2}{a}+
3\langle\!\langle f_1f_2^2\rangle\!\rangle
\frac{A_1}{a}
\left(\frac{A_2}{a}\right)^2
\right.
\nonumber\\
&&\phantom{a}-\left.
\langle\!\langle f_2^3\rangle\!\rangle
\left(\frac{A_2}{a}\right)^3
\right]+
15\left[
\langle\!\langle f_1^4\rangle\!\rangle
\left(\frac{A_1}{a}\right)^4-
4\langle\!\langle f_1^3f_2\rangle\!\rangle
\left(\frac{A_1}{a}\right)^3
\frac{A_2}{a}
\right.
\label{t13}\\
&&\phantom{a}\left.\left.+
6\langle\!\langle f_1^2f_2^2\rangle\!\rangle
\left(\frac{A_1}{a}\right)^2
\left(\frac{A_2}{a}\right)^2-
4\langle\!\langle f_1f_2^3\rangle\!\rangle
\frac{A_1}{a}
\left(\frac{A_2}{a}\right)^3
+\langle\!\langle f_2^4\rangle\!\rangle
\left(\frac{A_2}{a}\right)^4
\right]\right\}.
\nonumber
\ees
\noindent
Here the double angle brackets denote two successive
averaging procedures. The first one is the
averaging over the surface area of interacting
bodies. The second one is over all possible phase
shifts between the distortions situated on the
surfaces of interacting bodies against each other.
This second averaging is necessary because in the
experiment \cite{M-R} the measured Casimir force
was averaged over 26 scans (see Sec.~II).

Note that under condition (\ref{t12}) the result
(\ref{t13}) can be obtained in two ways: starting
from the Eqs.\,(\ref{t8}), (\ref{t9}) for a sphere
above a plate \cite{K-P} and applying  Force
Proximity Theorem \cite{FPT} to the Eq.\,(25) of
Ref.\cite{B-K-M} which is an analog of (\ref{t13})
for the configuration of two plane parallel plates.
As one would expect, the results coincide (in the
case of large-scale roughness violating the
condition (\ref{t12}) the special redefinition of
a distance is needed for the correct application of
Force Proximity Theorem \cite{B-K-R}).

\section{INVESTIGATION OF THE SURFACE ROUGHNESS}

Let us apply the result (\ref{t13}) to 
carefully calculate
the roughness corrections to the Casimir
force in the experiment \cite{M-R}. 
The roughness of the metal 
surface was measured with the same AFM. After the
Casimir force measurement the cantilever with
sphere was replaced with a standard cantilever
having a sharp tip. Regions of the metal plate
differing in size from $1\mum\times 1\mum$  to
$0.5\mum\times 0.5\mum$  were scanned with the AFM.
A typical surface scan is shown in Fig.~3.
The roughness of the sphere was investigated with
a SEM and found to be
similar to the flat plate. In the surface scan of
Fig.~3, the lighter tone corresponds to
larger height. 

As is seen from Fig.~3
the major distortions are the large separate
crystals situated irregularly on the surfaces. They
can be modeled approximately by the parallelepipeds
of two heights. As the analysis of several AFM
images shows, the height of highest distortions is
about $h_1=40\,$nm  and of the intermediate ones --- about
$h_2=20\,$nm. Almost all surface between the
distortions is covered by the stochastic roughness
of height $h_0=10\,$nm. It consists of small
crystals which are not clearly visible in Fig.3 due to the
vertical scale used.  All
together they form the  homogeneous background
of the averaged 
height $h_0/2$. The character of roughness 
on the plate and on the lens is quite similar.

Now it is possible to determine the height $H$
relative to which the middle value of the function,
describing the total roughness, is zero. It can be
found from the equation
\beq
(h_1-H)S_1+(h_2-H)S_2-
\left(H-\frac{h_0}{2}\right)S_0=0,
\label{t14}
\eeq
\noindent
where $S_{1,2,0}$ are, correspondingly, the surface
areas occupied by distortions of the heights 
$h_1$, $h_2$ and stochastic roughness.
Dividing (\ref{t14}) into the area of interacting 
surface 
$S=S_1+S_2+S_0$ one gets
\beq
(h_1-H)v_1+(h_2-H)v_2-
\left(H-\frac{h_0}{2}\right)v_0=0,
\label{t15}
\eeq
\noindent
where $v_{1,2,0}=S_{1,2,0}/S$ are the relative parts of
the surface occupied by the different kinds of
roughness. The analysis of the AFM pictures similar to
Fig.~3 gives us the values $v_1=0.11$, $v_2=0.25$,
$v_0=0.64$. Solving Eq.\,(\ref{t15}) we get the
height of the zero distortions level
$H=12.6\,$nm. The value of distortion amplitude
defined relatively to this level is
\beq
A=h_1-H=27.4\,\mbox{nm}.
\label{t16}
\eeq

Below two more parameters will also be used 
\beq
\beta_1=\frac{h_2-H}{A}\approx 0.231,
\qquad
\beta_2=\frac{H-h_0/2}{A}\approx 0.346.
\label{t17}
\eeq
\noindent
With the help of them the distortion function from
(\ref{t5}) can be represented as
\beq
f_1(x_1,y_1)=\left\{
\begin{array}{rcl}
1,&&(x_1,y_1)\in\Sigma_{S_1},\\
\beta_1,&&(x_1,y_1)\in\Sigma_{S_2},\\
-\beta_2,&&(x_1,y_1)\in\Sigma_{S_0},
\end{array}
\right.
\label{t18}
\eeq
\noindent
where $\Sigma_{S_1,S_2,S_0}$ are the regions of the
first interacting body surface occupied by the
different kinds of roughness.

The same representation is valid for $f_2$ also
\beq
f_2(x_2,y_2)=\left\{
\begin{array}{rcl}
-1,&&(x_2,y_2)\in\tilde\Sigma_{S_1},\\
-\beta_1,&&(x_2,y_2)\in\tilde\Sigma_{S_2},\\
\beta_2,&&(x_2,y_2)\in\tilde\Sigma_{S_0},
\end{array}
\right.
\label{t19}
\eeq
\noindent
$\tilde\Sigma_{S_1,S_2,S_0}$ are the regions of the
second interacting body surface occupied by the
distortions of
different kinds.

Note that the inequality (\ref{t12}) is easily satisfied. For 
the roughness under consideration
the characteristic lateral sizes of distortions are
$d_p,d_s\sim 200-300\,$nm as can be seen from Fig.~3.
At the same time $\sqrt{aR}> 3000\,$nm. Thus, Eq.\,(\ref{t13})
is applicable for the calculation of roughness corrections.

Now it is not difficult to calculate the coefficients
of expansion (\ref{t13}). One example is
\beq
\langle\!\langle f_1f_2\rangle\!\rangle=
-v_1^2-2\beta_1v_1v_2+2\beta_2v_1v_0
-\beta_1^2v_2^2+2\beta_1\beta_2 v_2v_0-
\beta_2^2 v_0^2=0,
\label{t20}
\eeq
\noindent
which follows from Eqs. (\ref{t15})--(\ref{t17}).
The results for the other coefficients are
\bes
&&
\langle\!\langle f_1^2\rangle\!\rangle =
\langle\!\langle f_2^2\rangle\!\rangle =
v_1+\beta_1^2 v_2+\beta_2^2 v_0,
\nonumber\\
&&
\langle\!\langle f_1^3\rangle\!\rangle =
-\langle\!\langle f_2^3\rangle\!\rangle =
v_1+\beta_1^3 v_2-\beta_2^3 v_0,
\quad
\langle\!\langle f_1f_2^2\rangle\!\rangle =
\langle\!\langle f_1^2 f_2\rangle\!\rangle =0,
\nonumber\\
&&
\langle\!\langle f_1^4\rangle\!\rangle =
\langle\!\langle f_2^4\rangle\!\rangle =
v_1+\beta_1^4 v_2+\beta_2^4 v_0,
\quad
\langle\!\langle f_1f_2^3\rangle\!\rangle =
\langle\!\langle f_1^3 f_2\rangle\!\rangle =0,
\nonumber\\
&&
\langle\!\langle f_1^2 f_2^2\rangle\!\rangle =
(v_1+\beta_1^2 v_2+\beta_2^2 v_0)^2.
\label{t21}
\ees

Substituting (\ref{t21}) into (\ref{t13}) we get the
final expression for the Casimir force with 
surface distortions included upto the fourth order
in relative distortion amplitude
\bes
&&
F_R(a)=F_0(a)
\left\{
\vphantom{\left[\left(\beta_2^2v_0\right)^2\right]
\frac{A^4}{a^4}}
1+12\left(v_1+\beta_1^2v_2+\beta_2^2v_0\right)
\frac{A^2}{a^2}\right.
\nonumber\\
&&\phantom{aaaaa}
+20\left(v_1+\beta_1^3v_2-\beta_2^3v_0\right)
\frac{A^3}{a^3}
\label{t22}\\
&&\phantom{aaaaa}\left.
+30\left[v_1+\beta_1^4v_2+\beta_2^4v_0+
3\left(v_1+\beta_1^2v_2+\beta_2^2v_0\right)^2\right]
\frac{A^4}{a^4}\right\}.
\nonumber
\ees

It should be noted that exactly the same result can be
obtained in a very simple way. To do this it is
enough to calculate the values of the
Casimir force (\ref{t11})
for six different distances which are possible
between the distorted surfaces, multiply them by the
 appropriate probabilities and then to summarize the
results
\bes
&&
F_R(a)=\sum\limits_{i=1}^{6}w_i F_0(a_i)
\equiv
v_1^2F_0(a-2A)
\label{t23}\\
&&\phantom{F(a)}
+2v_1v_2F_0\left(a-A(1+\beta_1)\right)
+
2v_2v_0F_0\left(a-A(\beta_1-\beta_2)\right)
\nonumber\\
&&\phantom{F(a)}+
v_0^2F_0(a+2A\beta_2)+
v_2^2F_0(a-2A\beta_1)+
2v_1v_0F_0\left(a-A(1-\beta_2)\right).
\nonumber
\ees
\noindent
The question arises of whether there is unique
definition of the distance $a$ between the
interacting bodies in Eqs.\,(\ref{t22}), (\ref{t23}).
This point is discussed in the next section in
connection with the reflectivity properties of the
metals covering the plate and the sphere.

\section{CORRECTIONS TO THE CASIMIR FORCE DUE TO
FINITE CONDUCTIVITY OF THE METALS}

The interacting bodies used in the experiment
\cite{M-R} were coated with 300\,nm of $Al$ in an
evaporator. The thickness of this metallic layer is
much larger than the penetration depth $\delta_0$
of electromagnetic oscillations into $Al$ for the
wavelengths (sphere-plate separations) of interest.
Taking $\lambda_p^{Al}= 100\,$nm as the 
approximative value
of the effective plasma wavelength of the electrons
in $Al$ \cite{Hbook} one gets
$\delta_0=\lambda_p^{Al}/(2\pi)\approx 16\,$nm. What
this means is the interacting bodies can be 
considered as made of $Al$ as a whole.
Although $Al$ reflects more than 90\% of the
incident electromagnetic oscillations in the
complete measurement range 
$100\,\mbox{nm}<\lambda <950\,$nm, 
some corrections to the Casimir force due to the
finiteness of its conductivity exist and should
be taken into account. In addition, to prevent the
oxidation processes, the surface of $Al$ in \cite{M-R}
was covered with $\Delta=20\,$nm layer of
$60\%Au/40\%Pd$. The reflectivity properties of this
alloy are much worse than of $Al$ (the effective
plasma wavelength of $Au$ is 
$\lambda_p^{Au}= 500\,$nm and the penetration
depth is $\tilde\delta_0\approx 80\,$nm).
Because of this, it is impossible to use 
Eq.\,(\ref{t11a})
to compare  theory and experiment
as it is only valid for ideal metals of infinite
conductivity. 
It is necessary
to take into account the finiteness of the metal
conductivity.

Let us start our discussion with the large distances
$a>\lambda_p^{Au}$ for which both $Al$ and
$Au/Pd$ are the good metals. In this case the
perturbation theory in the relative penetration depth
can be developed. This small parameter is the ratio
of an effective penetration depth $\delta_e$ 
(into both $Au/Pd$ and $Al$) and
a distance between the $Au/Pd$ layers $a$. The
quantity $\delta_e$, in its turn, is understood as
a depth for which the electromagnetic oscillations
are attenuated by a factor of $e$. It takes into account
both the properties of $Al$  and of $Au/Pd$ layers.
The value of $\delta_e$ can be found from the
equation
\beq
\frac{\Delta}{\tilde\delta_0}+
\frac{\delta_e-\Delta}{\delta_0}=1,
\qquad
\delta_e=\left(1-\frac{\Delta}{\tilde\delta_0}\right)
\delta_0+\Delta\approx 32\,\mbox{nm}.
\label{t25}
\eeq

The first order corrections to the Eq.\,(\ref{t11a})
was found in \cite{Harg,Schw} for the configuration
of two plane parallel plates. Together with the
second order correction found in \cite{M-T-p} the
result is
\beq
F_{\delta_e}(a)=F_0(a)\left(1-\frac{16}{3}
\frac{\delta_e}{a}+24\frac{\delta_e^2}{a^2}\right).
\label{t26}
\eeq

From the general expression for $F_{\delta_e}$ it is
seen that the Casimir force taking
into account the finite
conductivity is sign-constant for all $\delta_e$
and has a zero limit when $\delta_e\to\infty$. This
gives the possibility to obtain the simple
interpolation formula \cite{M-T-p}
\beq
F_{\delta_e}(a)\approx F_0(a)
\left(1+\frac{11}{3}\frac{\delta_e}{a}\right)^{-16/11}.
\label{t27}
\eeq
\noindent
From (\ref{t27}) we have the same result as in
(\ref{t26}) for small $\delta_e/a$, but it is
applicable in the wider range
$0\leq\delta_e/a\leq 0.2$.

Let us now expand (\ref{t27}) in powers of 
$\delta_e/a$ up to the fourth order inclusive and
modify the result by the use of Force Proximity 
Theorem \cite{FPT} to the case of a sphere above
a plate
\beq
F_{\delta_e}(a)=F_0(a)\left(1-4
\frac{\delta_e}{a}+
\frac{72}{5}\frac{\delta_e^2}{a^2}-
\frac{152}{3}\frac{\delta_e^3}{a^3}+
\frac{532}{3}\frac{\delta_e^4}{a^4}
\right).
\label{t28}
\eeq
\noindent
Here $F_0(a)$ is defined by (\ref{t11a}).

Now we combine both corrections --- 
one due to the surface
roughness and the second
due to the finite conductivity of the
metals. For this purpose we substitute the quantity
$F_{\delta_e}(a_i)$ from (\ref{t28})
into Eq.\,(\ref{t23}) instead of
$F_0(a_i)$. The result is
\beq
F(a)=\sum\limits_{i=1}^{6}
w_i F_{\delta_e}(a_i),
\label{t28a}
\eeq
\noindent
where different possible distances between the
surfaces with roughness and their probabilities were
introduced in (\ref{t23}). Eq.\,(\ref{t28a})
along with 
(\ref{t28}) describes the Casimir force between $Al$
bodies with $Au/Pd$ layers taking into account the
finite conductivity of the metals and surface
roughness for the distances
$a>\lambda_p^{Au}$.
Note that (\ref{t28a}) incorporates not only the
corrections to the surface roughness and finite
conductivity separately but also some ``crossed"
terms, i.e. the conductivity corrections to the
roughness ones.

Unfortunately, the Eq.\,(\ref{t28a}), strictly
speaking, cannot be used for the distances
$a<\lambda_p^{Au}$. The most rigorous
way of calculating the Casimir force in this range
is to apply the general Lifshitz theory without the
supposition that $a$ is much larger than the
characteristic absorption band of $Au/Pd$
 (this supposition
leads to the result (\ref{t11}) with a definition
(\ref{t3})). To do this the detailed information
is needed concerning the behaviour of the dielectric
permittivity of $Au/Pd$ on the imaginary frequency
axis. This information should reflect the absorption
bands of the alloy and the damping of free electrons
\cite{VB-O}. In doing so the actual dependence of the
Casimir force on $a$ could be calculated, where $a$
is the distance between the outer $Au/Pd$ layers.

At the same time, there exists a more simple,
phenomenological, approach to calculation of the
Casimir force for the distances less than the 
characteristic absorption wavelength 
of the $Au/Pd$ covering.
It uses the fact that the transmittance of 20\,nm
$Au/Pd$ films for the wavelength of around 300\,nm
is greater than 90\%. This transmission measurement
was made by taking the ratio of light transmitted
through a glass slide with and without the $Au/Pd$
coating in an optical spectrometer.

So high transmittance gives the possibility to
neglect the $Au/Pd$ layers when calculating the
Casimir force and to enlarge the distance between the
bodies by $2\Delta=40\,$nm when comparing the
theoretical and experimental results. With this
approach for the distances 
$a<\lambda_p^{Au}$, instead of
(\ref{t28a}), the following result is valid
\beq
F(a)=\sum\limits_{i=1}^{6}
w_i F_{\delta_0}(a_i+2\Delta),
\label{t29}
\eeq
\noindent
where the Casimir force with account of finite
conductivity is
defined by the Eq.\,(\ref{t28}).

\section{COMPARISON WITH THE EXPERIMENT}

Let us first consider large surface separations 
(the distance
between the $Au/Pd$ layers changes in the interval
610\,nm$\leq a\leq 910\,$nm). 
We  compare the results given by (\ref{t28a}) 
and (\ref{t29}) with experimental data.
In Fig.~4 the dashed
curve represents the results obtained by the 
Eq.\,(\ref{t28a}), and solid curve --- by the
Eq.\,(\ref{t29}). 
The experimental points are shown as open squares.
For eighty experimental points,
which belong to the range of $a$ under consideration,
the root mean square average deviation between 
theory and experiment in both cases is 
$\sigma=1.5\,$pN. It is notable that for the large
$a$ the same result is valid also if we use the
Casimir force from Eq.\,(\ref{t11a}) (i.e. without any
corrections) both for $a$ and for $a+2\Delta$. By this
is meant that for large $a$ the problem of the proper
definition of distance is not significant due to
the large scatter in experimental points due to the 
experimental uncertainty. The same
situation occurs with the corrections. At
$a+2\Delta=950\,$nm the correction due to roughness
(positive)
is of about 0.2\% of $F_0$ and the correction due
to finite conductivity (negative) is 6\% of $F_0$. 
Together they give the negative contribution 
which is also 6\%
of $F_0$. It is negligible if we take into account
 the relative error of force measurements 
at the extreme distance of 950\,nm
is approximately 660\% (this is because the Casimir
force is much less than the experimental uncertainty at
such distances).

Now we consider the range of smaller values of the
distance 80\,nm$\leq a\leq 460\,$nm (or, between
$Al$, 120\,nm$\leq a+2\Delta\leq 500\,$nm).
Here the Eq.\,(\ref{t29}) should be
used for the Casimir force. In Fig.~5 the Casimir
force $F_0(a+2\Delta)$ from (\ref{t11a}) is shown by
the dashed curve. The solid curve represents the
dependence calculated according to Eq.\,(\ref{t29}).
The open squares are the experimental points.
Taking into account all one hundred experimental
points belonging to the range of smaller distances
we get for the solid curve the value of the root mean
square deviation between theory and experiment
$\sigma_{100}=1.5\,$pN. If we consider more
narrow distance interval
80\,nm$\leq a\leq 200\,$nm which contains thirty
experimental points it turns out that 
$\sigma_{30}=1.6\,$pN
for the solid curve. In all the measurement range
80\,nm$\leq a\leq 910\,$nm the root mean square
deviation for the solid curves of Figs.~4,\,5 is
$\sigma_{223}=1.4\,$pN (223 experimental points). 
What this means is that 
the dependence
(\ref{t29}) gives equally good agreement with
experimental data in the region of small distances
(for the smallest ones 
the relative error of force measurement is about
1\%), in the region of large distances (where the
relative error is rather large) and in the whole
measurement range.  If one uses less sophisticated
expressions for the corrections to the Casimir force
due to the surface roughness and finite conductivity,
the value of $\sigma$ calculated for small $a$ would
be larger than in the whole range \cite{M-R}.

It is interesting to compare the obtained results with
those given by Eq.\,(\ref{t11a}), i.e. without
account of any corrections. In this case
for the interval
80\,nm$\leq a\leq 460\,$nm 
(one hundred experimental points) we have
$\sigma_{100}^0=8.7\,$pN. For the whole measurement
range
80\,nm$\leq a\leq 910\,$nm (223 points) there is
$\sigma_{223}^0=5.9\,$pN. 
It is evident that without 
appropriate treatment of the corrections to the Casimir
force the value of the root mean square deviation is
not only larger but also depends
significantly on the measurement range.

The comparative role of each correction is also quite
obvious. If we take into account only roughness
correction according to Eq.\,(\ref{t23}), then
one obtains
for the root mean square deviation in different
intervals: $\sigma_{30}^R=22.8\,$pN,
$\sigma_{100}^R=12.7\,$pN and 
$\sigma_{223}^R=8.5\,$pN. At $a+2\Delta=120nm$
the correction is 17\% of $F_0$. For the single 
finite conductivity correction calculated by
Eq.\,(\ref{t28}) with $\delta_0$ instead of
$\delta_e$ it follows:
$\sigma_{30}^{\delta}=5.2\,$pN,
$\sigma_{100}^{\delta}=3.1\,$pN and 
$\sigma_{223}^{\delta}=2.3\,$pN.
At 120\,nm this correction contributes\ --34\%
of $F_0$. (Note, that both corrections contribute\ 
--22\% of $F_0$ at 120\,nm, so that their
nonadditivity is demonstrated most clearly.)

Considering the case of small distances, we have
neglected the contribution of thin $Au/Pd$ layers
which are almost transparent for the essential
frequencies. The corrections to the Casimir force
due to these layers can be calculated considering 
them as being made of some effective dielectric with
a small permittivity $\varepsilon$ 
\cite{book}. Such corrections calculated with
$\varepsilon\approx 1.1$ lead, together with
(\ref{t29}), to the same value of $\sigma$ if we increase 
the values of all distances by 1\,nm.
If the effective value of permittivity would be
$\varepsilon\approx 1.2$ this is equivalent to 
addition of 3\,nm to all the distances without
changing of $\sigma$. As can well be imagined, the
corrections due to $Au/Pd$ layers are not essential
when it is considered that the absolute uncertainty
of distance measurements in the experiment
\cite{M-R} was about $\pm 5\,$nm.

\section{CONCLUSIONS AND DISCUSSION}

In the above, the surface roughness of the test bodies
used in the experiment \cite{M-R} on Casimir force
measurment was investigated with the use of AFM and
SEM. The corrections to this force due to both 
surface roughness and finite conductivity of the
metal were calculated up to the fourth order in
respective small parameters. The obtained theoretical
results for the Casimir force with both corrections
were confronted with the experimental data. 
The excellent agreement was demonstrated which is
characterized by almost the same value of the root
mean square deviation between theory and experiment
in the cases of small and large
space separations between the test bodies and in the complete
measurement range.

It was shown that the agreement between the theory 
and experiment is substantially worse if any one of 
the corrections is not taken into account. 
What this means is that the surface roughness
and finite conductivity corrections should be
taken into account in precision Casimir force measurements 
with space separations of
the order $1\mum$ and less. They will also be expected to 
play a strong role in experimental tests of the shape and 
topology dependences of the Casimir force. 

Further improvements in the precision can be achieved 
through the use of smoother metallic coatings, thinner
$Au$-layers (but of enough thickness to prevent the
oxidation processes of $Al$) and larger radius 
spheres to 
increase the values of force. The experimental uncertainties 
can be substantially reduced by use of 
lower temperatures to 
decrease the thermal noise in the AFM, 
and interferometric 
detection of cantilever deflection \cite{Rugar}. 
This will
provide the opportunity to increase the accuracy of
Casimir force measurements and to obtain more strong
constraints on the constants of hypothetical
long-range interactions and light elementary
particles. Such constraints for the different ranges
of Compton wavelengths of hypothetical particles
were already obtained in \cite{BGKM1} from
the experiment \cite{Lam} and in \cite{BGKM2} from
the experiment \cite{M-R}. There is reason to hope
that within the next few years the Casimir effect
will become a strong competitor to the more
traditional physical phenomena which can provide us
with new data about long-range interactions and 
light elementary particles.

\section*{ACKNOWLEDGMENTS}
    
G.L.K. and V.M.M are grateful
to the Physics Department of the Federal University
of Para\'{\i}ba, where this work was partly
done, for their kind
hos\-pi\-ta\-li\-ty.

\newpage
\begin{center} {\Large {\bf List of captions}} \end{center}
\begin{tabular}{lcp{142mm}}
& &\\
{\bf Fig.1.}& & Schematic diagram of the 
experimental setup. Application of voltage to the
piezo results in the movement of the plate towards
the sphere.\\          
{\bf Fig.2.}& & A typical force curve as a function
of the distance moved by the plate.
 \\
{\bf Fig.3.}& & A typical Atomic Force
Microscope scan of the metal surface. 
The lighter tone corresponds to larger height as 
shown by the bar graph
on the left.
\\
{\bf Fig.4.}& & The measured average Casimir force 
for large distances as a function of plate-sphere
separation is shown as open squares. The theoretical
Casimir force with corrections to surface roughness and
finite conductivity is shown by the solid line (when
the space separation is defined as the distance
between $Al$ layers) and by the dashed line (with the
distance between $Au/Pd$ layers). 
\\
{\bf Fig.5.}& & The measured average Casimir force
for small distances as a function of plate-sphere
separation is shown as open squares. The theoretical Casimir force 
with corrections to 
surface roughness and finite
conductivity is shown by the solid line, and without
any correction -- by the dashed line.

\end{tabular}

\end{document}